\documentclass[lettersize,journal]{IEEEtran}
\usepackage{tabularx}
\usepackage{algorithm}
\usepackage{algpseudocode}
\usepackage{amsmath,amssymb,amsfonts}
\usepackage{array}
\usepackage{multirow}
\usepackage{booktabs}
\usepackage{xspace}
\usepackage{enumitem}
\usepackage{listings}
\usepackage{xcolor}
\usepackage{dblfloatfix}
\usepackage{afterpage}
\usepackage{tikz}
\usetikzlibrary{shapes,arrows,positioning,fit}
\usepackage{url}
\usepackage{cite}
\usepackage[hidelinks]{hyperref}
\usepackage{placeins}

\makeatletter
\lst@AddToHook{PreInit}{\vspace{0.35em}}
\makeatother

\newcommand{\toolname}[0]{\textsc{ORION}\xspace}

\begin{document}

\title{Behavioral Regression Detection for Pull Requests via Structured Oracles}
\author{Yiwen~Fu,
        Binhang~Qi,
        Xiao~Li,
        Runlin~Liu,
        Hailong~Sun,
        and~Xiang~Gao%
\thanks{Yiwen Fu, Xiao Li, and Runlin Liu are with
Beihang University, Beijing, China
(e-mail: fuyiwen@buaa.edu.cn;
axiao@buaa.edu.cn;
runlin22@buaa.edu.cn).}%
\thanks{Binhang Qi is with the National University of Singapore,
Singapore, and also with Beihang University, Beijing, China
(e-mail: qibh@nus.edu.sg).}%
\thanks{Xiang Gao and Hailong Sun are with
Beihang University, Beijing, China, and also with
the Hangzhou Innovation Institute of Beihang University,
Hangzhou, China
(e-mail: xiang\_gao@buaa.edu.cn; sunhl@buaa.edu.cn).}%
\thanks{Corresponding authors: Binhang Qi; Xiang Gao.}%
}
\maketitle

\begin{abstract}
Software regressions refer to unintended behavioral changes introduced by code modifications and remain difficult to detect at the pull-request (PR) level. Existing approaches either rely on incomplete tests or directly analyze PR contents with LLMs or coding agents, lacking structured behavioral reasoning and explicit behavioral oracles.
We present \toolname, a staged framework for PR-level regression detection. \toolname first extracts PR intent and identifies impacted functions through AST-based impact region analysis.
It then synthesizes function-level structured behavioral oracles that encode preserved and changed behaviors, setup requirements, and control-flow hints as explicit constraints derived from PR intent, static risk signals, and code changes in impacted functions.
Guided by these oracles, \toolname generates executable tests and performs oracle-coverage-guided augmentation and static repair. Finally, it differentially executes the same tests on pre- and post-change versions and classifies a PR as regression-inducing or not using deterministic rules and constrained LLM reasoning.
To enable rigorous evaluation, we construct RegBench-Py, the first benchmark for PR-level regression detection in Python, containing 154 real-world PRs with verified labels. \toolname achieves F1-scores
of 0.6125 with DeepSeek-Chat-V3.2 and 0.5578 with Claude-4.5-haiku, the highest among the methods evaluated on the full benchmark.
Compared with Testora, \toolname improves F1-score by
0.2542--0.2882 across the two LLM backends, and it also outperforms the Codex and Claude Code agentic baselines in F1-score. A scope-aware comparison with PatchGuru further shows that neither method consistently dominates on the common single-function subset, while \toolname supports a broader range of multi-function PRs. Applying \toolname to 1,459 recently merged PRs further identifies
9 reproducible bug candidates, 8 of which have been confirmed by project maintainers. \toolname requires approximately 15 minutes per PR, demonstrating its practical applicability.
\end{abstract}


\section{Introduction}
\label{sec:intro}

Software regression poses a fundamental threat to software reliability, which is unintended behavioral change introduced by code changes, e.g., in the form of pull requests (PRs).
As software systems undergo frequent changes to fix defects, extend features, optimize efficiency, and enhance maintainability, such code changes can unintentionally undermine existing correct behaviors and introduce unforeseen functional flaws ~\cite{leung1990study, jamil2016software}.
Such regression issues are widespread in real-world development and often stay undetected throughout regular development and review processes, as even in well-tested software, the available test suite may not cover the code modified by a code change.

To address this problem, a broad spectrum of regression testing techniques has been developed.  
Symbolic execution and hybrid analysis have been explored to enhance test coverage over modified code for systematically exploring program paths ~\cite{cadar2008klee, majumdar2007hybrid}.
Meanwhile, search-based approaches have also been applied to both maximize structural coverage ~\cite{fraser2011evosuite} and systematically reveal behavioral differences between program versions ~\cite{harman2012search}. 
More recently, learning-based methods, particularly those leveraging large language models (LLMs), have emerged to generate regression tests directly from code changes and associated artifacts (e.g., PR descriptions), enabling a more flexible and context-aware form of test generation ~\cite{pradel2025testora}. PatchGuru~\cite{lepatchguru} further infers patch oracles from natural-language development artifacts and operationalizes them as executable assertions with concrete inputs.

However, existing approaches still have significant limitations.
\textbf{First, behavioral oracles for regression detection are under-specified.}
Many existing approaches either reason directly from PR intent and code changes or rely on weak or implicit test oracles, without constructing a
structured intermediate representation that explicitly separates behaviors
that should be preserved from behaviors that are intended to evolve.
Without such behavioral boundaries, it remains difficult to determine what
properties tests should validate and whether observed differences reflect
intended changes or actual regressions.
This long-standing test oracle problem hinders precise and reliable regression detection ~\cite{6963470, pezze2014automated}.
\textbf{Second, fine-grained structural changes are often ignored.} Automated test generation techniques based on symbolic execution and search-based methods can explore program paths, but fail to detect regressions from fine-grained structural changes such as parameter removal, default value shifts, exception drift, silent \texttt{**kwargs} argument swallowing, and function-to-generator conversion. These changes often do not break control flow or trigger explicit failures, but instead cause silent behavioral divergence that requires deep structural and compatibility reasoning—something these traditional techniques cannot provide ~\cite{brito2020you, kula2018empirical, huang2023scalable}.
\textbf{Third, deeper behavioral implications are often overlooked.} By deeper implications, we refer to behavioral consequences that are not directly visible from changed lines alone, but emerge when the change is interpreted under PR intent, preserved behavioral contracts, caller assumptions, and control-flow conditions. For example, changing a default argument may be intended for new calls but may silently alter legacy call behavior.
Recent learning-based approaches, particularly those leveraging LLMs or coding agents, often directly reason over PR descriptions and code diffs to generate tests or assess regression risks.
This tends to bias the analysis toward surface-level changes, common usage scenarios, and simple execution paths, while overlooking deeper behavioral implications~\cite{ryan2024code}.
As a result, critical regression-triggering conditions are often insufficiently covered ~\cite{ryan2024code, siddiq2024using}.

To address these limitations, we propose \toolname (Oracle-guided Regression Detection), a framework for behavioral regression detection in pull requests based on structured oracles. It models the task as a staged reasoning-and-execution pipeline that separates behavioral understanding from test synthesis. First, it extracts PR intent and identifies impacted functions through AST-based impact region analysis. It then constructs function-level \textbf{structured behavioral oracles} that encode preserved behaviors, changed behaviors, setup requirements, and control-flow hints. Guided by these oracles, the framework generates tests to cover behavioral boundaries and executes them on pre- and post-change versions. Finally, regression decisions are made via a hybrid strategy that combines deterministic rules with constrained LLM reasoning to distinguish intended behavioral changes from actual regressions.

To enable rigorous evaluation, we construct RegBench-Py, the first benchmark explicitly focused on PR-level regression detection in Python projects.
Unlike existing datasets \cite{just2014defects4j, song2022regminer, widyasari2020bugsinpy}, RegBench-Py adopts a reproducible labeling
procedure that links regression-inducing PRs to explicit developer evidence, filters out samples with unresolved build dependencies or non-reproducible local builds, and validates clean PRs through repository-history screening and independent manual reinspection.
The final dataset contains balanced regression and non-regression PRs (154 in total), providing a reliable foundation for our evaluation and future research.

Experimental results show that \toolname achieves F1-scores of 0.6125 with DeepSeek-Chat-V3.2 and 0.5578 with Claude-4.5-haiku; both are
higher than the F1-scores of the baselines evaluated on the full RegBench-Py benchmark. Compared with Testora, \toolname improves F1-score by 0.2882 and 0.2542 under the two
LLM backends, respectively. Claude Code with Claude Sonnet 5 achieves an F1-score of 0.5246, providing a stronger agentic baseline than Codex. We further conduct a scope-aware comparison
with PatchGuru~\cite{lepatchguru}; neither method consistently dominates on the commonly completed single-function PRs, while \toolname supports a broader range of multi-function PRs.
Finally, in an extended real-world evaluation over 1,459 recently merged PRs, \toolname identifies nine reproducible bug candidates,
eight of which have been confirmed by project maintainers.
These results demonstrate the effectiveness and practical value of \toolname.

In summary, our contributions are as follows:
\begin{itemize}[leftmargin=*]
    \item We propose \toolname, a staged framework for PR-level regression detection that integrates structured oracle construction, oracle-guided test generation, and hybrid regression classification.
    \item We release RegBench-Py, the first benchmark specifically designed for PR-level regression detection in Python. RegBench-Py consists of 77 regression and 77 non-regression PRs from GitHub projects.
   The regression-inducing PRs are supported by explicit developer evidence and manually verified for reproducibility. The clean PRs are screened using subsequent repository history, and all 77 selected clean PRs are independently re-inspected by two annotators using PR discussions, linked issues, developer tests, and subsequent repository history.
    \item We conduct an extensive evaluation against five baselines, including Testora, PatchGuru, direct LLM reasoning, Codex, and Claude Code. On the full RegBench-Py benchmark, \toolname achieves an F1-score of up to 0.6125, while a separate scope-aware comparison with PatchGuru highlights its broader applicability to multi-function PRs. In an extended real-world study of 1,459 PRs, \toolname identifies nine reproducible bug candidates, eight of which have been confirmed by maintainers.
\end{itemize}

\newcommand{\entity}[3]{\textbf{#2} (\textit{#3})}

\section{Motivating Example}

\begin{figure*}[t]
\centering
\includegraphics[width=1\textwidth]{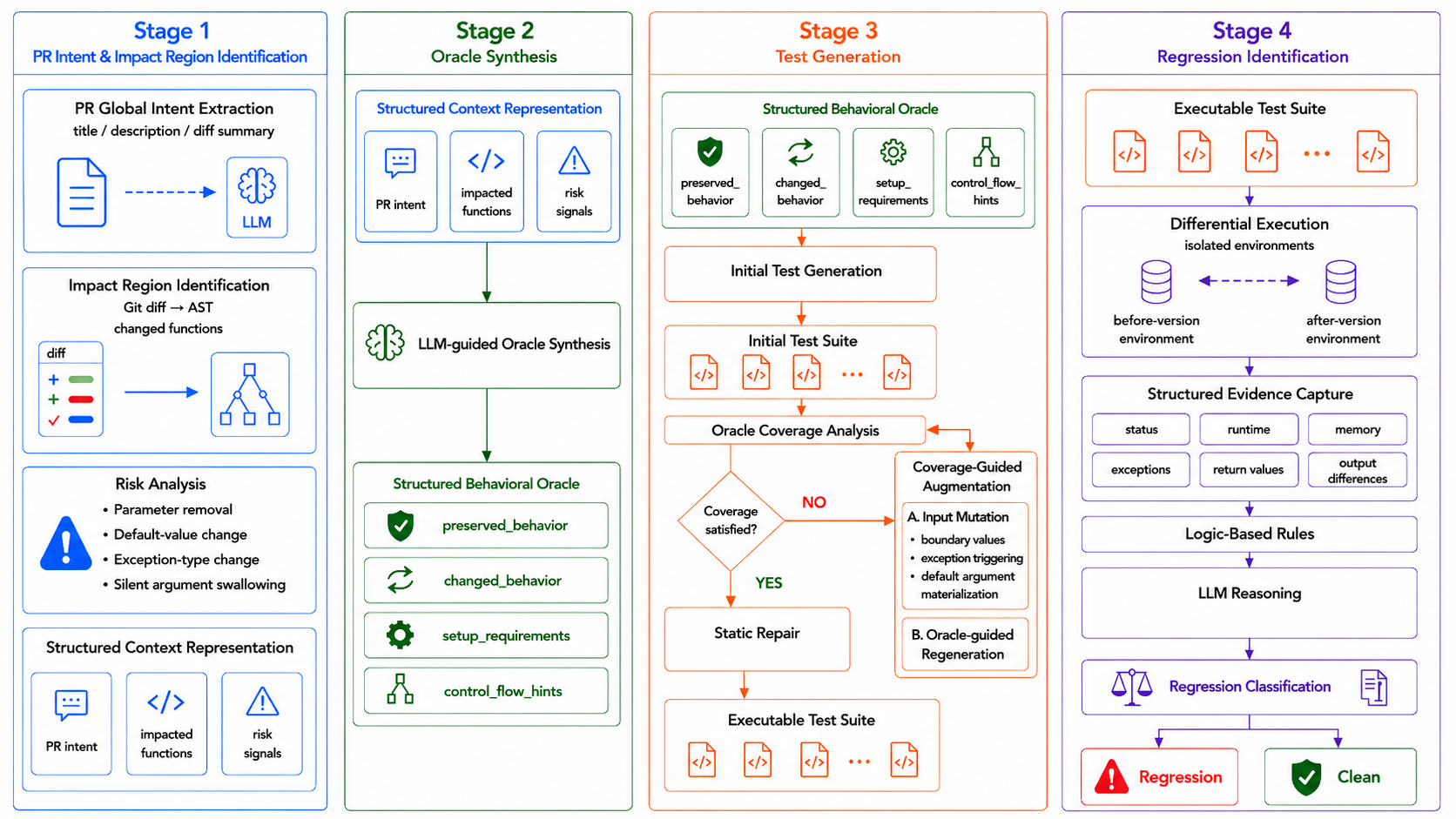}
\caption{Overview of \toolname. }
\label{fig:pipeline}
\end{figure*}

Listing~\ref{pr} presents a representative example, which illustrates the limitations of the state-of-the-art approach.
This example is a real-world regression introduced by PR \#9613 
of project \textit{yt-dlp} ~\cite{ytdlp_pr_9613}, a video downloader supporting downloading videos from Vimeo, Patreon, etc. 
Listing~\ref{pr} shows the message and code changes of this PR, which aims to fix a bug when downloading Vimeo videos embedded in Patreon post.

The pre-PR version simply processes any URL (lines 5-10) without validating whether the URL is valid.
When provided with an invalid URL, an ``HTTP Error 404'' will be triggered.
To fix this bug, the patch tries to check whether the URL is valid (lines 12-13) and only proceeds valid URLs (line 14).
Function \texttt{\_request\_webpage()} handles general-purpose HTTP requests that take the first argument as a normal HTTPS URL.
However, the output of function \texttt{\_smuggle\_referrer()} (line 11) is a specialized struct that can only be handled by VimeoIE class, and \texttt{\_request\_webpage()} will always return false.
As a result, a regression was introduced since previously valid URLs will be identified as invalid, causing normal behaviors to be broken.

\begin{lstlisting}[language=Python, caption=Code change of yt-dlp PR \#9613, label=pr]
!\textbf{PR Message:}! [patreon] fix some vimeo embed downloads. This change adds checks to verify if external embeds of Patreon posts are still valid before trying to download them, and if it is invalid then it tries other options...

!\textbf{Changed Code:}!
if v_url:
-    return {
-        **info,
-        '_type': 'url_transparent',
-        'url': VimeoIE._smuggle_referrer(v_url, 'https://patreon.com'),
-        'ie_key': 'Vimeo',
-    }
+    v_url = VimeoIE._smuggle_referrer(v_url, 'https://patreon.com')
+    if self._request_webpage(v_url, video_id,
+        'Checking Vimeo embed URL', fatal=False, errnote=False):
+        return self.url_result(v_url, VimeoIE, url_transparent=True, **info)
\end{lstlisting}

Existing tools, e.g., Testora~\cite{pradel2025testora}, only analyze code changes and focus on the superficial intent of the PR, such as ``fix invalid Vimeo embed''. 
Thus, they only synthesize test cases for the new logic: checking whether invalid embeds can be correctly handled.
As a result, it fails to detect this regression.

\begin{lstlisting}[language=, label=oracle, caption=Partial of inferred behavioral oracle.]
"preserved_behavior": [
  "Posts with working vimeo embeds should extract successfully as before"]
"changed_behavior": [
  "Posts with invalid vimeo embeds should now succeed by falling back to post_file"
]
\end{lstlisting}

In contrast, \toolname first generates a function-level structured behavioral oracle, as shown in Listing~\ref{oracle}.
Guided by this oracle, \toolname generates a test case that constructs a Patreon post with a working Vimeo embed, asserting that extraction succeeds without error.
The test passes before the PR, but fails after the PR with the error: ```NoneType' object has no attribute \texttt{\_first\_webpage\_request}''. 
This exception is captured during differential execution and finally classified as a regression by \toolname's LLM-assisted assessment.

\section{Methodology}
\label{sec:methodology}

Our approach detects regressions by transforming a PR into a staged reasoning and execution pipeline. As shown in Figure~\ref{fig:pipeline}, instead of directly generating tests from diffs, \toolname consists of four stages:

\noindent \textbf{Stage 1 (PR Intent and Impact Region Identification):} \toolname extracts PR intent, maps diff changes to affected functions via AST parsing, and produces structured impact regions with explicit risk signals. 

\noindent \textbf{Stage 2 (Oracle Synthesis):} \toolname synthesizes a structured oracle from PR intent and static risk signals, defining preserved\_behavior invariants, changed\_behavior evolution scenarios, setup requirements, and control-flow hints. 

\noindent \textbf{Stage 3 (Test Generation):} \toolname generates executable tests under oracle guidance, covering behavioral boundaries and control-flow
changes.

\noindent \textbf{Stage 4 (Regression Identification):} \toolname executes the tests on pre- and post-change versions and classifies behavioral differences using Logic-based rules and LLM reasoning.

\subsection{PR Intent and Impact Region Identification}
Oracle generation requires both a macroscopic understanding of developer intent and a microscopic understanding of structural code changes. Raw diffs alone are insufficient: they intermingle semantic modifications with syntactic noise, and they do not explicitly encode compatibility expectations. Therefore, before synthesizing oracles, \toolname constructs a structured contextual representation that integrates PR-level intent with function-level impact analysis. This representation serves as the primary input to the oracle synthesis stage.

\subsubsection{PR Global Intent Extraction}
To distinguish intended behavior evolution from unintended regressions, \toolname extracts the global intent of the PR from its metadata, including title, description, and diff summary. The extracted intent is normalized into four fields: change\_type, expected\_behavior\_change, backward\_compatibility, and high\_level\_expectations. Specifically, change\_type provides a coarse prior for distinguishing intended behavior evolution from behaviors that should remain preserved. expected\_behavior\_change defines the intended scope of behavioral modification and determines the boundary within which preserved-behavior constraints are enforced. backward\_compatibility captures declared breaking changes, removed features or parameters, and migration paths, allowing the oracle to distinguish documented interface evolution from suspicious compatibility drift. high\_level\_expectations summarizes developer-stated goals into structured constraints that anchor oracle reasoning to the PR's declared objectives.

\subsubsection{Impact Region Identification}
While global intent provides macroscopic expectations, regressions originate from concrete modified code units. \toolname therefore identifies the minimal set of directly impacted functions and extracts structural deltas that may introduce behavioral drift.

\toolname parses the Git diff to obtain precise line ranges of modifications and maps them to the abstract syntax tree (AST) of each affected file. Functions, methods, constructors, or class attributes containing modified lines are marked as directly changed functions. For each such function, \toolname retrieves the pre- and post-change implementations and records deterministic structural deltas, including removed parameters, changed default values, exception-type changes, possible silent argument swallowing through \texttt{**kwargs}, and conversion from a regular function to a generator. These patterns are selected because prior studies identify parameter removal and default-value modification as common sources of breaking changes~\cite{brito2020you,kula2018empirical}, while Python's dynamic features can make such compatibility drift difficult to detect before execution~\cite{huang2023scalable,fan2023large}. \toolname uses these lightweight checks to expose subtle compatibility hazards without introducing heavyweight analyses such as data-flow or taint analysis.

Beyond directly changed functions, \toolname constructs a lightweight relational context containing immediate callers, callees, and sibling methods, along with a minimal call graph representation. This contextual information is preserved separately and later used during test generation to ensure executable call paths, but oracle synthesis remains scoped to directly impacted functions.

The output of this stage integrates PR intent, localized structural deltas, and lightweight relational context into a compact, machine-readable representation. By excluding unrelated code while exposing compatibility-sensitive risk signals, it provides a task-focused input for oracle synthesis rather than raw diffs.

\subsection{Oracle Synthesis}

Directly feeding diff content, PR intent, and static analysis results into an LLM to generate tests is feasible in principle, yet exhibits practical limitations. The model must simultaneously perform behavior-risk reasoning and executable test construction, which increases cognitive load and may cause fine-grained structural risks to be overlooked. Moreover, static constraints are difficult to reliably preserve within unconstrained generation. To address this issue, \toolname introduces an intermediate oracle layer between diff analysis and test generation. The oracle serves as a function-level behavioral contract that decouples logical reasoning from code synthesis. 
It specifies behavioral boundaries without producing executable code, while the test generation phase implements tests based on this oracle. This layered design improves generation stability and provides interpretable behavioral coverage boundaries.

Oracle construction is not applied indiscriminately to all modified functions. Instead, it targets functions with meaningful behavioral impact. The following cases are excluded:
(1) modifications limited to comments, formatting, or documentation;
(2) non-executable or abstract interface functions;
(3) trivial forwarding or duplicated wrapper functions;
(4) cases where AST parsing or oracle construction fails.
When the number of impacted functions exceeds the bounded analysis
budget, ORION applies a budget-based ranking rather than treating
some functions as semantically unimportant. Functions are ranked
according to PR-intent relevance, local call connectivity, function
size/type, and control-flow complexity, and the highest-ranked
functions are selected for oracle synthesis. Thus, ``low priority''
denotes relative priority under a bounded analysis budget rather than
a judgment that a function has no significant behavioral impact.
This filtering mechanism ensures that oracle synthesis focuses on semantically relevant changes rather than mechanically covering all modifications.

Oracle generation consumes the static risk signals produced in Stage 1. When a risk signal is present, it is injected into the oracle-generation prompt as a standardized requirement, encouraging the LLM to encode the corresponding preservation or changed-behavior constraint. In the absence of such signals, oracle synthesis relies on PR intent and semantic reasoning over the before/after function implementations.

The final oracle is represented as a structured JSON behavioral oracle:

\begin{lstlisting}[language=,caption={Function-level oracle schema}]
{
  "function_id": "<string>",
  "code_diff_summary": ["<string>"],
  "setup_requirements": ["<execution prerequisite>"],
  "preserved_behavior": ["<behavior that must not change>"],
  "changed_behavior": ["<expected behavior change>"],
  "behavior_notes": ["<additional notes or status flags>"],
  "control_flow_hints": {
    "new_paths": ["<new or modified path>"],
    "important_conditions": ["<key decision conditions>"]
  }
}
\end{lstlisting}

The field function\_id uniquely identifies the modified function and remains consistent with the impact region analysis stage. code\_diff\_summary summarizes the core semantic change introduced by the diff. setup\_requirements specifies environmental and dependency prerequisites to ensure test executability. preserved\_behavior defines behaviors that should remain stable and therefore provides the primary boundary for identifying potential regressions, whereas changed\_behavior characterizes intended behavioral evolution to mitigate false positives. behavior\_notes records auxiliary annotations or status flags. Finally, control\_flow\_hints preserves structural guidance: new\_paths highlights newly introduced or modified control-flow branches, and important\_conditions enumerates decision predicates critical for distinguishing behavioral changes.

By explicitly encoding behavioral invariants, expected evolution, and control-flow context, the oracle provides an operational intermediate representation between change analysis and test synthesis, enabling a principled separation between reasoning and implementation.

\subsection{Test Generation}

A generated test is an executable unit that validates the behavioral constraints defined within the oracle. Test generation therefore aims not only to produce runnable code, but also to cover critical behavioral boundaries and newly introduced control-flow paths under strict executability requirements.
To achieve this, \toolname adopts a three-stage pipeline consisting of generation, coverage augmentation, and static repair.

\subsubsection{Oracle-Guided Generation}

Test generation is driven exclusively by the oracle rather than directly by the raw diff. The system extracts verification targets from preserved\_behavior, changed\_behavior, and new\_paths, and constructs prompts that explicitly encode these behavioral requirements. The lightweight relational context from Stage 1 is also used to construct feasible invocation paths when direct function calls are insufficient. The LLM is therefore constrained to instantiate executable scenarios consistent with predefined behavioral oracles instead of re-inferring change semantics.

To control test explosion, \toolname caps initial generation at 24 base tests per PR, a practical engineering budget for limiting generation costs on large PRs rather than an empirically established optimal threshold.
Within this budget, \toolname prioritizes high-impact functions using a simple heuristic score:
\[
Priority(f) = \alpha \cdot |CB(f)| + \beta \cdot |PB(f)|.
\]
Here, $CB(f)$ and $PB(f)$ denote the numbers of changed\_behavior and preserved\_behavior entries for function $f$, respectively.
The parameters $\alpha$ and $\beta$ control the relative contributions of these two types of oracle entries to function prioritization.
We use fixed heuristic weights of $\alpha=5$ and $\beta=2$ to give greater initial priority to functions with more intended-change requirements while retaining preservation requirements in the allocation.
These values are engineering choices rather than empirically established optima.
The weighting controls initial test allocation rather than the importance of either type of behavior for regression classification.
Uncovered oracle entries of both types remain eligible for the subsequent oracle-coverage-guided augmentation stage.

Each generated test must return structured JSON and explicitly include an oracle\_ref field linking the test to a specific oracle region and index, enabling deterministic coverage tracking.

\subsubsection{Coverage-Guided Augmentation}

After base generation, the system performs structural coverage analysis using the oracle\_ref indices rather than textual matching. For each function, \toolname constructs a mapping from oracle region types to covered index sets and compares them against the full set of oracle entries. Uncovered preserved\_behavior, changed\_behavior, or new\_paths items are recorded as coverage gaps.

Enhancement proceeds under a global augmentation cap of 40 tests.
This budget exceeds the base-test cap because augmentation targets behavioral constraints left uncovered by the initial suite, allowing additional tests to address identified oracle-coverage gaps.
Two complementary strategies guide this process.
First, input mutation extends existing tests by perturbing boundary values, explicitly triggering exceptional branches, materializing default arguments, or consuming generator outputs while preserving the original testing intent.
Second, oracle-guided regeneration incorporates uncovered oracle entries into the prompt and requires the LLM to generate tests that reference the corresponding \texttt{oracle\_ref}.
Functions are prioritized by the number of uncovered oracle items so that the augmentation budget is directed toward the remaining behavioral constraints.

\subsubsection{Static Repair for Executability}

Because LLM-generated tests frequently contain syntactic and import-related defects, \toolname applies a purely static repair phase to guarantee executability without altering test semantics. This phase performs Markdown code block extraction and indentation normalization, identifier typo correction while protecting language keywords and string literals, symbol resolution against import mappings derived from the source code, automatic insertion of missing imports, degradation of hallucinated submodule imports to verified root-package imports, removal of unused imports, and final AST-level syntax validation in a single deterministic repair pass.

Importantly, \toolname deliberately avoids semantic adaptation of tests to either the pre- or post-change versions of the program. Regression detection relies on executing identical tests across program versions; any logic-level modification could mask genuine behavioral divergence or introduce inequivalence between versions. Therefore, repair is restricted to minimal executability guarantees and never modifies the logical intent of generated tests.

\subsection{Regression Identification}

After test generation, \toolname executes each test in isolated environments to determine whether observed behavioral changes indicate a regression. This process follows a staged pipeline: isolated execution $\rightarrow$ structured objective data capture $\rightarrow$ rule-based classification $\rightarrow$ LLM-assisted reasoning for unresolved cases. Such a design mitigates limitations of traditional regression detection, including subjective judgments and non-reproducible decisions, by prioritizing objective signals and deterministic reasoning~\cite{yoo2012regression,huang2023scalable}.

Tests are run independently in Conda/Docker environments corresponding to the \textit{before} and \textit{after} program versions. Both environments are fully consistent in dependencies, system configuration, and resource limits, with the only difference being the code under test. During execution, \toolname captures objective metrics, including execution status, runtime, memory usage, exceptions, return values, and output differences. These data are stored as structured JSON and provide the objective execution evidence for regression classification.

Regression decisions follow a deterministic-first strategy. \toolname first applies hard rules to clear-cut cases before invoking the LLM. Identical before/after executions are labeled as \texttt{clean}. Infrastructure or setup failures, such as missing dependencies, environment errors, or resource-limit violations unrelated to the code change, are filtered out before classification and are not used as regression evidence. Tests targeting newly introduced functions without a meaningful before-version baseline are treated as non-regression evidence unless they also expose failures in existing preserved call paths. Severe performance degradation is labeled as \texttt{regression} when the after-version runtime is at least twice the before-version runtime and the slowdown is not attributable to infrastructure noise or an explicitly intended performance tradeoff.

For unresolved cases, \toolname invokes an LLM constrained by structured execution data, PR intent, and oracle-derived behavioral constraints. The LLM does not freely judge regressions from raw outputs; instead, it answers five standardized yes/no questions:
$Q_1$: does the test expose an observable before/after behavioral difference?
$Q_2$: does the status change indicate a failure that is not covered by changed\_behavior?
$Q_3$: does the exception difference violate preserved\_behavior?
$Q_4$: does the return value or stdout difference violate preserved\_behavior rather than reflect intended evolution?
$Q_5$: does the runtime increase exceed the slowdown threshold without being an intended performance tradeoff?

The final label is derived from the structured answers according to the following consistency rule:

\[
\mathrm{regression\_label} =
\begin{cases}
\text{regression} & \text{if } Q_1=\text{yes} \land \exists Q_{2\text{--}5}=\text{yes}, \\
\text{clean} & \text{otherwise}.
\end{cases}
\]

ORION first classifies individual executable cases. At the PR level,
a PR is classified as regression-inducing if at least one case is
classified as regression; otherwise, it is classified as clean.
Infrastructure-failure cases do not themselves trigger a regression,
and PRs containing only such cases are additionally flagged for
human review.
This formulation keeps the decision grounded in objective execution evidence while using the LLM only for semantic alignment between observed differences, PR intent, and oracle constraints. By combining isolated differential execution, deterministic rules, and constrained LLM reasoning, \toolname provides reproducible and explainable regression detection.
\section{RegBench-Py: A Benchmark for PR-level Regression Detection}
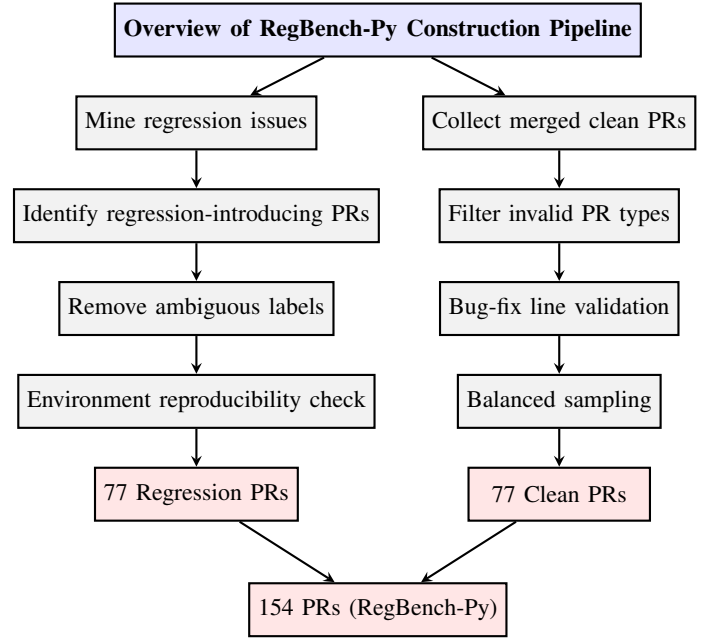
\begin{figure}[t]
  \centering
  \begin{tikzpicture}[
    node distance=0.5cm and 0.8cm, 
    block/.style={rectangle, draw, thick, text centered, minimum height=0.7cm, minimum width=2.2cm, fill=gray!10, font=\small},
    startblock/.style={rectangle, draw, thick, text centered, minimum height=0.7cm, minimum width=2.2cm, fill=blue!10, font=\small},
    endblock/.style={rectangle, draw, thick, text centered, minimum height=0.7cm, minimum width=2.4cm, fill=red!10, font=\small},
    arrow/.style={thick, -stealth}
  ]

  \node (title) [startblock] {\textbf{Overview of RegBench-Py Construction Pipeline}};

  \node (pos1) [block, below=of title, xshift=-2.4cm] {Mine regression issues};
  \node (pos2) [block, below=of pos1] {Identify regression-introducing PRs};
  \node (pos3) [block, below=of pos2] {Remove ambiguous labels};
  \node (pos4) [block, below=of pos3] {Environment reproducibility check};
  \node (pos_end) [endblock, below=of pos4] {77 Regression PRs};

  \node (neg1) [block, below=of title, xshift=2.4cm] {Collect merged clean PRs};
  \node (neg2) [block, below=of neg1] {Filter invalid PR types};
  \node (neg3) [block, below=of neg2] {Bug-fix line validation};
  \node (neg4) [block, below=of neg3] {Balanced sampling};
  \node (neg_end) [endblock, below=of neg4] {77 Clean PRs};

  \node (final) [endblock, below=of pos_end, xshift=2.4cm, yshift=-0.3cm] {154 PRs (RegBench-Py)};

  \draw [arrow] (title) -- (pos1);
  \draw [arrow] (pos1) -- (pos2);
  \draw [arrow] (pos2) -- (pos3);
  \draw [arrow] (pos3) -- (pos4);
  \draw [arrow] (pos4) -- (pos_end);

  \draw [arrow] (title) -- (neg1);
  \draw [arrow] (neg1) -- (neg2);
  \draw [arrow] (neg2) -- (neg3);
  \draw [arrow] (neg3) -- (neg4);
  \draw [arrow] (neg4) -- (neg_end);

  \draw [arrow] (pos_end) -- (final);
  \draw [arrow] (neg_end) -- (final);

  \end{tikzpicture}
  \caption{Pipeline of RegBench-Py Dataset Construction}
  \label{fig:regbenchpy_pipeline}
\end{figure}

To address the lack of high-quality and reproducible benchmarks for Python regression detection, we construct RegBench-Py, the first benchmark specifically designed for PR-level regression detection in Python projects. 
Existing datasets used in prior work (e.g., Testora) rely on scattered regression samples without standardized construction procedures, which limits reproducibility and fair evaluation. 
RegBench-Py is designed to ensure (1) \textit{authenticity}, as all samples originate from real-world pull requests, (2) \textit{reproducibility}, requiring each regression sample to be buildable in an isolated environment, and (3) \textit{unambiguous labeling}, where regression introduction must be explicitly confirmed by developers. 
To ensure balanced evaluation, the dataset contains equal numbers of regression and non-regression pull requests. 
All samples are collected from 207 active Python projects on GitHub with at least 100 stars and commits within the last 12 months. The final RegBench-Py benchmark contains 154 pull requests, including 77 regression-introducing PRs and 77 clean PRs, spanning 65 unique repositories. Figure~\ref{fig:regbenchpy_pipeline} illustrates the overall construction pipeline.

\begin{figure}[t]
\centering
\includegraphics[width=1\linewidth]{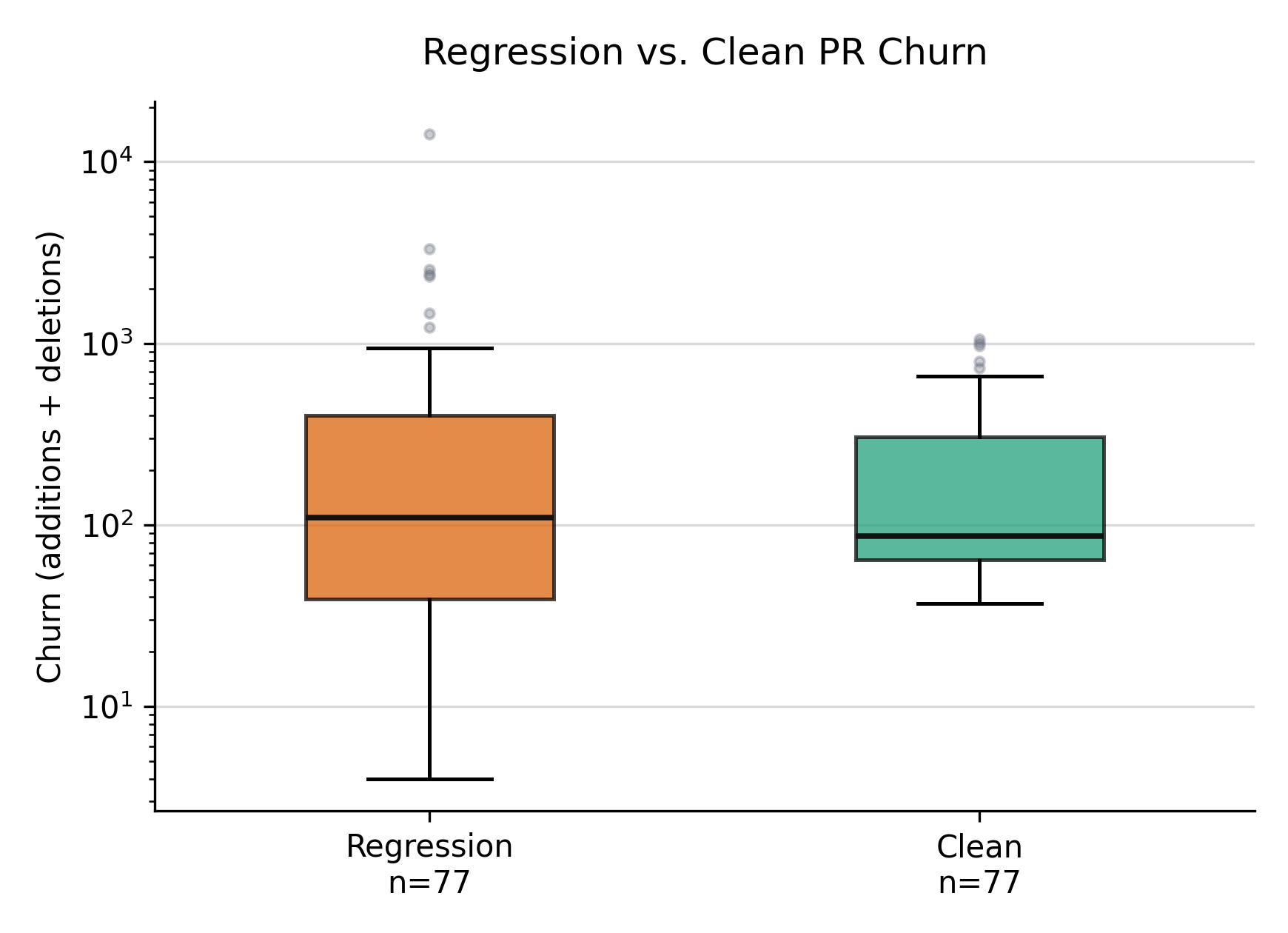}
\caption{Distribution of PR churn in RegBench-Py on a logarithmic scale. Churn is computed as the sum of added and deleted lines. Boxes indicate interquartile ranges, center lines indicate medians, whiskers show non-outlier ranges, and points denote outlier PRs.}
\label{fig:regbenchpy_churn}
\end{figure}

\textbf{Positive samples (regression-introducing PRs).}
A positive sample is defined as a pull request that explicitly introduces a regression bug and can be successfully built in an isolated environment. 
As shown in Figure~\ref{fig:regbenchpy_pipeline}, the construction consists of four main stages.
First, we mine regression-related issues using the query \texttt{label:regression is:closed language:python}, obtaining 2,465 issues related to code logic regressions while excluding documentation or configuration regressions. 
Second, we identify regression-introducing PRs by matching developer statements in issue descriptions and comments using regular expressions (e.g., ``introduced by \#PR'', ``regression from \#PR''), yielding 186 candidate PRs. 
To avoid ambiguous labels, we remove samples containing uncertain expressions (e.g., ``guess'', ``suspect'', ``may''), resulting in 101 candidates. 
Third, we verify environment reproducibility by constructing isolated Docker/Conda environments for each candidate and checking whether the corresponding project version can be successfully built. 
We discard 24 PRs due to build failures caused by issues such as dependency compilation errors (e.g., C/C++/Rust/Cython extensions), incompatibility with modern toolchains, missing installation files (e.g., \texttt{setup.py}, \texttt{pyproject.toml}), submodule retrieval failures, or missing commits. 
Finally, two researchers independently inspect the remaining samples to confirm that the regression attribution is explicitly supported by developer discussions, resulting in 77 regression-introducing PRs.

\textbf{Negative samples (clean PRs).}
Negative samples represent pull requests that do not introduce regressions.
We define a clean PR as one whose modified lines are not subsequently changed by any identified bug-fixing PR within 12 months after merging and for which manual review finds no concrete evidence of an introduced regression.
As shown in Figure~\ref{fig:regbenchpy_pipeline}, we first scan all PRs across the 207 projects and retain merged PRs with at least a 12-month post-merge observation window, source-code changes beyond documentation, configuration, and tests, and modified lines that can be extracted using PyDriller.
Next, we identify subsequent bug-fixing PRs using labels (e.g., \texttt{bug}, \texttt{fix}, \texttt{regression}, \texttt{patch}), textual indicators (e.g., ``fix bug'', ``crash'', ``revert'', ``regression''), or issue-closing keywords (e.g., ``fix'', ``resolve'', ``close'').
If any such PR modifies lines previously changed by a candidate within the observation window, the candidate is removed.
We then randomly sample 77 PRs from the remaining candidates to balance the dataset.
To validate these negative labels, two authors independently re-inspected all 77 selected PRs using PR discussions, linked issues, developer tests, and subsequent repository history.
They initially agreed on 75/77 PRs (97.4\%); the two disagreements were jointly adjudicated, and both PRs were retained only after both annotators found no concrete evidence of an introduced regression.
These screening and review procedures establish \emph{clean} as an operational negative label rather than proof of absolute regression absence.

Figure~\ref{fig:regbenchpy_churn} shows the churn distribution of regression and clean PRs. The typical churn of the two groups is comparable, while regression PRs show a wider spread and a heavier tail. This suggests that RegBench-Py includes both moderate-size regression-introducing changes and a small number of large regression PRs, reflecting the skewed nature of real-world PR complexity.

\section{Experiment}
To evaluate \toolname, we address the four research questions:
\begin{itemize}[leftmargin=*]
    \item \textbf{RQ1: Classification accuracy on RegBench-Py.} 
     How accurately can \toolname detect regression-inducing pull requests on the RegBench-Py benchmark, and how does it compare with the five evaluated baselines?
     \item \textbf{RQ2: Impact of key components.}  How do the major components of \toolname affect end-to-end regression detection performance?
    \item \textbf{RQ3: Average runtime and cost overhead.} 
    What are the average time and LLM cost overheads for processing a single PR?
    \item \textbf{RQ4: Effectiveness on real-world regressions.} 
    How effective is our approach at finding unknown regressions in newly merged PRs from real projects?
\end{itemize}

\subsection{Experimental Setup}

\subsubsection{Baselines and Configuration}
We compare our approach with five baselines.
The first is Testora, an existing tool specifically designed for regression detection at the pull-request level.
We evaluate Testora under two
configurations: an adapted setting that removes its native filters for full-benchmark evaluation, and its original native setting for a scope-faithful comparison.

The second baseline is PatchGuru, a closely related PR-level regression detection approach.
PatchGuru's current implementation requires a PR diff to modify exactly one source function.
Therefore, we first identify the \textit{RegBench-Py} PRs satisfying this requirement and conduct a direct comparison on the common subset for which both PatchGuru and \toolname produce valid outputs.

The third baseline is a simple LLM baseline that directly judges regression risks using raw PR metadata and diff content.
This baseline receives the same PR information used by our approach, including the PR title, description, code diff, and comments, and is prompted to directly determine whether the PR introduces a regression.

The fourth baseline is Codex, an agentic programming baseline built on OpenAI's GPT-5-series Codex programming agent.
For each PR, Codex receives the same PR-level information, including the title, description, code diff, and comments, and is asked to analyze the change and determine whether it introduces a regression.

The fifth baseline is Claude Code with Claude Sonnet 5, which serves as another agentic programming baseline.
Like Codex, it receives the same PR-level information and is asked to analyze the change and determine whether it introduces a regression.
Claude Code is evaluated on the full \textit{RegBench-Py} benchmark.

For LLM-based analysis, we evaluate \toolname, Testora, and the simple LLM baseline with DeepSeek-Chat-V3.2 and Claude-4.5-haiku to assess cross-model robustness.
Codex and Claude Code are evaluated separately as agentic baselines because they use their respective agent backends rather than the two LLM backends used by the other methods.
All experiments use consistent API parameters, prompt settings, and execution environments where applicable.

\subsubsection{Metrics}
\label{sec:metrics}

To quantitatively evaluate the performance of our approach and baselines under various LLM backends, we first clarify the standard confusion matrix entries tailored to regression PR detection:
\begin{itemize}[leftmargin=*]
    \item TP: Correctly identified regression-inducing PRs.
    \item TN: Correctly identified non-regression PRs.
    \item FP: Non-regression PRs misclassified as regression.
    \item FN: Regression-inducing PRs misclassified as non-regression.
\end{itemize}
Then we calculate Precision, Recall, and F1-score.

\subsubsection{Implementation}
  Our prototype is implemented in Python 3 and consists of approximately 10,104 lines of code. The system performs repository analysis, PR change extraction, and regression detection through a multi-stage pipeline. 

For static code analysis, we use Python's built-in \texttt{AST} module to parse source files and perform lightweight analyses such as import validation and symbol resolution.
Generated tests are validated using AST-based checks to ensure correct imports and executable code structure.

Target projects are installed in isolated Conda/Dockers environments, with separate environments created for the pre-commit and post-commit versions of each PR. All experiments are conducted on a server running Ubuntu~22.04.

\subsection{RQ1: Classification Accuracy on RegBench-Py}
\label{subsec:rq1}

This research question evaluates the ability of \toolname to identify regression-inducing pull requests (PRs) through PR-level binary classification on \textit{RegBench-Py}.
We compare \toolname with five baselines: Testora, PatchGuru, a direct LLM prompting baseline, Codex, and Claude Code.
\toolname, Testora, PatchGuru, and the direct LLM baseline are evaluated with DeepSeek-Chat-V3.2 and Claude-4.5-haiku to assess cross-model robustness.
Codex uses its OpenAI GPT-5-series Codex agent backend, while Claude Code uses Claude Sonnet 5.
All executable methods run in the same Conda/Docker environments.
We adopt Precision, Recall, and F1-score to quantify classification performance, with detailed definitions provided in Section~\ref{sec:metrics}.

We evaluate Testora under both adapted and native filtering settings.
For full-benchmark evaluation, the adapted setting removes its original 200-line diff limit and PR-level filters that exclude PRs with more than three non-test files, no non-test source files, documentation-only changes, or comment-only changes.
To assess performance under its original configuration, we also evaluate Testora with these filters enabled and compare it with \toolname on the resulting subset.
Testora's multi-label outputs are mapped to binary predictions: \texttt{regression} is mapped to Regression, while \texttt{intended\_change}, \texttt{coincidental\_fix}, and \texttt{checked} are mapped to Clean.

Under the adapted setting, \toolname and the direct LLM baseline successfully process all 154 PRs, whereas Testora produces valid outputs for 152 PRs and fails on the remaining two.
These failures are primarily caused by extreme diff characteristics, including excessively long diffs that exceed the model's effective reasoning capacity and overly short diffs that lead to uninformative responses.
We count these failures as incorrect predictions, contributing a false positive for a clean PR or a false negative for a regression-inducing PR, to account for end-to-end failures.
PatchGuru is evaluated separately because its implementation requires each PR diff to modify exactly one source function.
We first identify PRs satisfying this requirement and then compare PatchGuru with \toolname on the common subset for which both methods produce valid outputs.
We report PatchGuru's applicability and execution completion separately from its classification performance on this subset.

\textbf{Classification results.}
Table~\ref{tab:regbench} summarizes the full-benchmark results.
\toolname achieves the highest Recall and F1-score under both shared LLM backends.
With DeepSeek, it obtains a Recall of 0.6364 and an F1-score of 0.6125; with Claude, these values are 0.5325 and 0.5578, respectively.
The direct LLM baseline achieves higher Precision than \toolname under both backends, reaching 0.7368 with DeepSeek and 0.6333 with Claude.
However, its Recall is only 0.1818 and 0.2468, respectively, indicating that direct judgment from raw PR information misses many regression-inducing PRs.
Testora achieves F1-scores of 0.3243 with DeepSeek and 0.3036 with Claude, showing limited effectiveness on \textit{RegBench-Py}.

Among the agentic baselines, Codex achieves a Precision of 0.5833, a Recall of 0.1818, and an F1-score of 0.2772.
Claude Code achieves a Precision of 0.7111, a Recall of 0.4156, and an F1-score of 0.5246.
Although Claude Code performs better than Codex, both configurations of \toolname achieve higher Recall and F1-score.
Overall, \toolname provides a more balanced precision--recall trade-off on the full benchmark, improving Recall while maintaining competitive Precision.
Nevertheless, the remaining false positives and false negatives suggest that \toolname is best used as a non-blocking CI assistant or a tool for analyzing high-risk PRs, with developers reviewing its findings.

\begin{table}[t]
\centering
\caption{Classification performance on all 154 PRs in RegBench-Py.
Testora uses the adapted filtering setting. Bold values indicate
the best results within each shared-backend group.}
\label{tab:regbench}
\resizebox{\linewidth}{!}{%
\begin{tabular}{llccc}
\toprule
\textbf{Method} & \textbf{Backend} &
\textbf{Precision} & \textbf{Recall} & \textbf{F1} \\
\midrule
Codex & OpenAI Codex agent & 0.5833 & 0.1818 & 0.2772 \\
Claude Code & Claude Sonnet 5 & 0.7111 & 0.4156 & 0.5246 \\
\midrule
Direct LLM & DeepSeek & \textbf{0.7368} & 0.1818 & 0.2917 \\
Testora & DeepSeek & 0.5294 & 0.2338 & 0.3243 \\
\toolname & DeepSeek & 0.5904 & \textbf{0.6364} & \textbf{0.6125} \\
\midrule
Direct LLM & Claude & \textbf{0.6333} & 0.2468 & 0.3551 \\
Testora & Claude & 0.4857 & 0.2208 & 0.3036 \\
\toolname & Claude & 0.5857 & \textbf{0.5325} & \textbf{0.5578} \\
\bottomrule
\end{tabular}%
}
\end{table}

Testora's native filters retain 94 of the 154 PRs, including 41 regression-inducing and 53 clean PRs.
Table~\ref{tab:testora_native} compares Testora and \toolname on this same subset.
With DeepSeek, Testora achieves a Precision of 0.3000, a Recall of 0.1463, and an F1-score of 0.1967, whereas \toolname achieves 0.4884, 0.5122, and 0.5000, respectively.
With Claude, Testora achieves 0.2632, 0.1220, and 0.1667, compared with 0.4571, 0.3902, and 0.4211 for \toolname.
Thus, \toolname's advantage over Testora persists when the comparison is restricted to PRs accepted by Testora's original filters.

\begin{table}[t]
\centering
\caption{Classification performance on the 94 PRs retained by
Testora's native filters: 41 regression-inducing and 53 clean PRs.}
\label{tab:testora_native}
\small 
\setlength{\tabcolsep}{4pt}
\begin{tabular}{llccc}
\toprule
\textbf{Method} & \textbf{Backend} &
\textbf{Precision} & \textbf{Recall} & \textbf{F1} \\
\midrule
Testora & DeepSeek & 0.3000 & 0.1463 & 0.1967 \\
\toolname & DeepSeek & \textbf{0.4884} & \textbf{0.5122} & \textbf{0.5000} \\
\midrule
Testora & Claude & 0.2632 & 0.1220 & 0.1667 \\
\toolname & Claude & \textbf{0.4571} & \textbf{0.3902} & \textbf{0.4211} \\
\bottomrule
\end{tabular}%
\end{table}

For PatchGuru, 48 of the 154 PRs (31.2\%) satisfy its single-function requirement.
PatchGuru successfully completes 33 of these PRs, including 12 regression-inducing and 21 clean PRs.
The remaining 15 runs fail mainly because the generated driver does not exercise the target function or the generated specification does not terminate cleanly.
Table~\ref{tab:patchguru_common} reports the confusion counts for PatchGuru and \toolname on the 33 common completed PRs.
With DeepSeek, both methods identify two regression-inducing PRs, while PatchGuru produces five false positives and \toolname produces six.
With Claude, PatchGuru identifies one regression-inducing PR with three false positives, whereas \toolname identifies two with seven false positives.
Neither method consistently dominates on this restricted subset.
Accordingly, these results do not establish a classification advantage for \toolname over PatchGuru on common completed cases.
The main distinctions of \toolname are its broader applicability to multi-function PRs and its separation of behavioral specification from executable-test instantiation.

\begin{table}[t]
\centering
\caption{Classification results on the 33 PRs completed by both
PatchGuru and \toolname: 12 regression-inducing and 21 clean PRs.}
\label{tab:patchguru_common}
\small 
\setlength{\tabcolsep}{4pt}
\begin{tabular}{llrrrr}
\toprule
\textbf{Method} & \textbf{Backend} &
\textbf{TP} & \textbf{FP} & \textbf{FN} & \textbf{TN} \\
\midrule
PatchGuru & DeepSeek & 2 & 5 & 10 & 16 \\
\toolname & DeepSeek & 2 & 6 & 10 & 15 \\
\midrule
PatchGuru & Claude & 1 & 3 & 11 & 18 \\
\toolname & Claude & 2 & 7 & 10 & 14 \\
\bottomrule
\end{tabular}%
\end{table}

\textbf{Qualitative analysis.}
We examine the behavioral constraints involved in the 49 true-positive PRs from the reported DeepSeek run.
Of these, 25 involve both preservation and intended-change constraints, 18 involve intended-change constraints only, and six involve preservation constraints only.
The corresponding prediction groups have Precisions of 0.500, 0.818, and 0.545, respectively.
We also examine the five explicit structural-risk patterns and find that only 13 of the 154 PRs contain at least one such pattern.
The remaining 141 PRs are assessed through \toolname's general LLM-based behavioral judgment over pre- and post-change execution evidence.

To examine whether the detected regressions correspond to those reported by developers, two authors independently compare the generated tests and execution evidence with the developer-reported regression.
A PR is counted as consistent only when both annotators judge the evidence to be fully or partially related to the reported regression; disagreements are treated conservatively.
This analysis identifies 38 consistent cases among the 49 true positives.

\textbf{Run-to-run variability.}
We assess LLM nondeterminism through three DeepSeek runs.
The full-benchmark results in Table~\ref{tab:regbench} come from the first run rather than a best-of-multiple selection.
Across the three runs, the standard deviations of Precision, Recall, and F1-score are 0.040, 0.057, and 0.048, respectively, indicating moderate run-to-run variability.

\subsection{RQ2: Impact of Key Components}
\label{subsec:rq2}

To understand the contribution of each major stage in \toolname, we conduct a controlled ablation study on RegBench-Py with DeepSeek-Chat-V3.2 as the backbone LLM. 
This study follows the staged design of \toolname: analysis-guided oracle generation, oracle-guided test generation, oracle-driven augmentation, and intent/oracle-aware final judgment. The goal is to examine where structured oracles contribute most to end-to-end regression detection. 

We construct four ablated variants by removing or weakening one key stage at a time:

\begin{itemize}[leftmargin=*]
    \item \textbf{w/o analysis-guided oracle generation}: 
    Oracles are synthesized from the raw PR diff without static-analysis-derived risk signals.
    \item \textbf{w/o oracle-guided test generation}: 
    Tests are generated directly from the PR diff, bypassing intent extraction, impact analysis, oracle synthesis, and target-view construction.
    \item \textbf{w/o oracle-driven augmentation}: 
    Only the initially generated tests are used, without oracle coverage checking or oracle-driven augmentation.
    \item \textbf{w/o intent/oracle context in final judgment}: 
    The final judgment retains the deterministic rules and five-question protocol, but removes PR intent and oracle context from the LLM input.
    \item \textbf{Full \toolname}: 
    The complete pipeline described in Section~\ref{sec:methodology}.
\end{itemize}

\par\noindent\textbf{Experimental Setup.}
All variants are evaluated on the full RegBench-Py dataset under identical execution settings. We report Precision, Recall, and F1-score to measure end-to-end binary classification performance. The results are shown in Table~\ref{tab:ablation}.

\begin{table}[t]
\centering
\caption{Ablation study results on RegBench-Py.}
\label{tab:ablation}
\resizebox{\linewidth}{!}{%
\begin{tabular}{lccc}
\toprule
\textbf{Setting} & \textbf{Precision} & \textbf{Recall} & \textbf{F1} \\
\midrule
w/o analysis-guided oracle generation & 0.5161 & 0.6234 & 0.5647 \\
w/o oracle-guided test generation & 0.4444 & 0.2597 & 0.3279 \\
w/o oracle-driven augmentation & 0.5441 & 0.4805 & 0.5103 \\
w/o intent/oracle context in final judgment & 0.5875 & 0.6104 & 0.5987 \\
Full \toolname & \textbf{0.5904} & \textbf{0.6364} & \textbf{0.6125} \\
\bottomrule
\end{tabular}
}
\end{table}

As shown in Table~\ref{tab:ablation}, the full \toolname achieves the best overall performance. Removing oracle-guided test generation causes the largest degradation, reducing Recall from 0.6364 to 0.2597 and F1-score from 0.6125 to 0.3279. This indicates that the structured oracle is important for guiding the LLM toward regression-relevant test scenarios instead of relying only on raw PR diffs.

Oracle-driven augmentation also has a clear impact. Without this stage, Recall drops to 0.4805 and F1-score drops to 0.5103, showing that the initial tests alone leave some oracle-specified behaviors uncovered. Removing intent/oracle context from the final judgment leads to a smaller decrease, with Recall dropping to 0.6104 and F1-score to 0.5987. These results show that structured oracles contribute across the pipeline, with the largest impact appearing in test generation and augmentation.

\subsection{RQ3: Average Runtime and Cost Overhead}
\label{subsec:runtime_cost}
\begin{figure}[t]  
    \centering
    \includegraphics[width=0.95\linewidth]{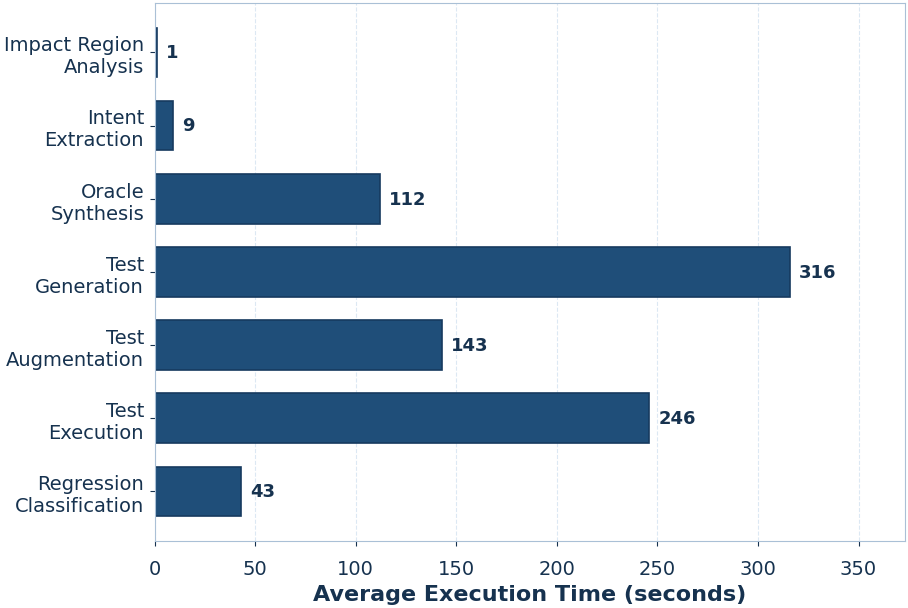}
    \caption{Runtime breakdown of \textsc{\toolname} pipeline.}
    \label{fig:time_breakdown}
\end{figure}

We analyze the runtime, token consumption, and approximate API cost of \toolname to assess its practical overhead. On average, \toolname takes approximately 870 seconds ($\approx$15 minutes) per PR. As shown in Figure~\ref{fig:time_breakdown}, the overhead is dominated by LLM-enabled stages and environment provisioning, while lightweight static analysis contributes negligibly. Test generation is the most expensive LLM stage, and test execution is another major bottleneck because most of its time is spent on Conda/Docker environment setup.

\begin{figure}[t]
    \centering
    \includegraphics[width=0.95\linewidth]{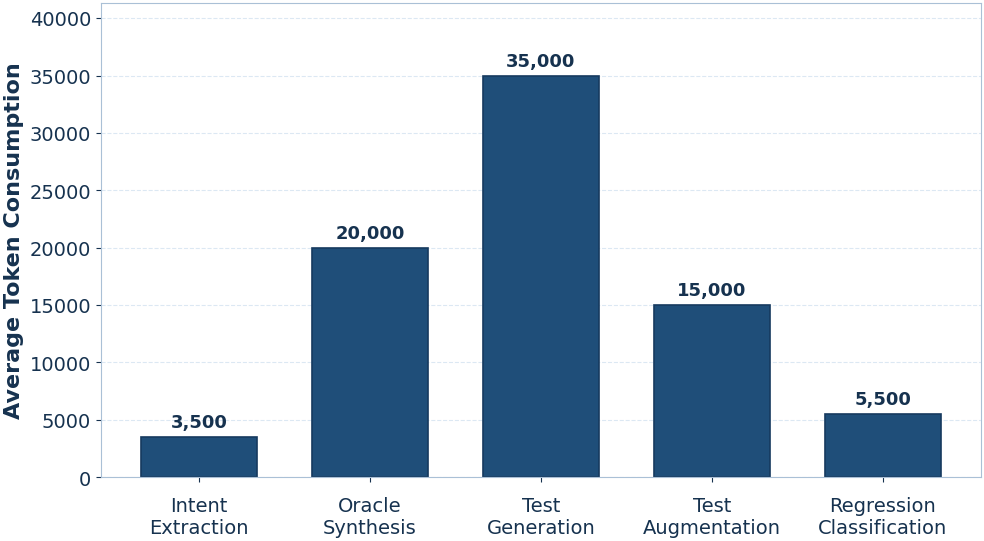}
    \caption{LLM token consumption per PR.}
    \label{fig:token_consumption}
\end{figure}

For LLM cost, each PR consumes approximately 60K--100K tokens across intent extraction, oracle synthesis, test generation, augmentation, and final classification, as shown in Figure~\ref{fig:token_consumption}. This corresponds to about \$0.01--\$0.02 per PR with DeepSeek-Chat and \$0.12--\$0.23 per PR with Claude Haiku 4.5. Overall, monetary cost is modest, whereas runtime remains the main practical bottleneck; parallel execution and environment caching can further reduce this overhead.

\subsection{RQ4: Effectiveness on Real-World Regressions}
Finally, we evaluate whether \toolname can discover unknown, unreported regressions
in real pull requests from popular open-source projects.
Our goal in this study is to assess the practical effectiveness of \toolname in realistic development settings, rather than to perform large-scale mining.

We selected four widely used Python projects from diverse domains: FastAPI for web/API frameworks, SymPy for symbolic mathematics, NetworkX for graph analysis, and Keras for deep learning. For each project, we collected PRs merged after June 1, 2025 using the query \texttt{is:pr is:merged merged:>=2025-06-01}. We retained PRs with valid base branches and 1--10 modified Python files, excluding test-only, documentation-only, formatting, typing, versioning, backport, and CI-only changes. After filtering, \toolname was applied to 1,459 PRs in total: 198 from NetworkX, 352 from SymPy, 127 from FastAPI, and 782 from Keras, as summarized in Table~\ref{tab:rq4_subjects}.
Across the complete study, ORION produces 28 warnings. When
revisiting these warnings, 9 are no longer reproducible on the
current repository versions because later upstream changes removed
the corresponding behavior. We manually inspect the remaining 19
currently reproducible warnings and identify 9 reproducible bug
candidates. Eight have been confirmed by project maintainers, while
one remains under discussion.

\begin{table}[t]
\centering
\caption{Subjects and PRs evaluated in the real-world study.}
\label{tab:rq4_subjects}
\setlength{\tabcolsep}{4pt}
\begin{tabular}{@{} l l r r @{}}
\toprule
\textbf{Project} & \textbf{Domain} & \textbf{Approx. Stars} & \textbf{\#PRs} \\
\midrule
NetworkX & Graph analysis & 17.3k & 198 \\
SymPy & Symbolic mathematics & 15.0k & 352 \\
FastAPI & Web/API framework & 103.3k & 127 \\
Keras & Deep learning framework & 64.4k & 782 \\
\midrule
Total & -- & -- & 1,459 \\
\bottomrule
\end{tabular}
\end{table}

In total, \toolname identified 9 reproducible, previously unreported
bug candidates in these recent real-world PRs. Among them, 8 have
been confirmed by maintainers and 1 remains under discussion, as shown in Table~\ref{tab:real_world_regressions}. These findings cover several forms of behavioral regression introduced by code evolution, including boundary-condition errors, backward-compatibility breaks, semantic logic changes, error-handling drift, and runtime exceptions caused by unsafe evaluation or incorrect field access. 

\begin{table}[t]
\centering
\caption{Real-world regressions detected by \toolname.}
\label{tab:real_world_regressions}
\setlength{\tabcolsep}{3pt} 
\begin{tabular}{@{} l l p{1.5cm} p{3.2cm} l @{}}
\toprule
\textbf{ID} & \textbf{Project} & \textbf{Type} & \textbf{Regression Description} & \textbf{Status} \\
\midrule
1 & NetworkX & Boundary & Similarity function raises exception for small graphs without auto-clamping parameters. & Confirmed \\
2 & NetworkX & Error handling & Invalid backend parameter triggers ImportError instead of being stored as an attribute. & Confirmed \\
3 & NetworkX & Numerical stability & HITS normalization divides by zero after an eigensolver change, producing infinite or NaN scores. & Confirmed \\
4 & NetworkX & Semantic logic & VF2++ isomorphism functions return reversed node mappings, breaking the original mapping direction. & Confirmed \\
5 & NetworkX & Semantic logic & Positive-weight self-loops overwrite zero self-distances in Floyd--Warshall, producing incorrect shortest-path distances. & Confirmed \\
6 & FastAPI & Compatibility & Public attribute removed without backward compatibility or deprecation. & Confirmed \\
7 & FastAPI & Runtime exception & NameError in parsing stringified type hints under Python 3.10+ due to unsafe evaluation. & Confirmed \\
8 & Keras & Runtime exception & AttributeError in ZIP extraction caused by wrong file name field access. & Confirmed \\
\cmidrule{1-5}
9 & SymPy & Semantic logic & Removed imaginary value checks, failing to return EmptySet for complex-real intersections. & Discussing \\

\bottomrule
\end{tabular}
\end{table}

Overall, this study shows that \toolname can be applied to hundreds of recently merged PRs across multiple active Python ecosystems. The goal of RQ4 is not to estimate the prevalence of regressions in these projects, but to evaluate whether \toolname can surface concrete, previously unreported behavioral regressions under a reproducible PR selection protocol. The results show that structured-oracle-guided regression detection can uncover non-trivial real-world issues that evade existing quality assurance processes.

\section{Threats to Validity}
\label{sec:threats}

\par\noindent\textbf{Internal validity:}
\toolname reconstructs pre- and post-change environments with Conda/Docker, where underspecified dependencies, unavailable packages, or toolchain incompatibilities may cause execution failures unrelated to regressions.
It also relies on LLM-generated oracles and PR intent modeling, so implicit developer intent or inaccurate oracles may affect test generation and final classification.
\toolname is more effective for regressions with explicit observable differences, such as return values, exceptions, CLI/config/IO, and serialization behavior.
It may miss cases involving complex states, remote services, caches, asynchronous behavior, or optional dependencies, and may produce false positives for intended API evolution or over-sensitive exception/message differences.
The average 15-minute runtime makes \toolname more suitable for asynchronous CI/CD analysis or high-risk PR screening than interactive review.
LLM-based stages are nondeterministic and may produce different oracles, tests, or final judgments across repeated runs.
The repeated DeepSeek-V3.2 evaluation in RQ1 shows moderate run-to-run variability, which should be considered when interpreting single-run performance.
Moreover, because the exact training corpora of the evaluated LLMs are unavailable, we cannot rule out their prior exposure to public GitHub PRs included in \textit{RegBench-Py}.
Although \toolname requires executable pre- and post-change evidence, this requirement does not eliminate the potential threat of training-data leakage.

\par\noindent\textbf{External validity:}
Our evaluation is based on RegBench-Py and real-world Python projects, so the results may not generalize to other languages, ecosystems, or development workflows.
RegBench-Py is limited in size and may not cover all regression patterns.
Although regression and clean PRs have comparable typical churn, regression PRs show a wider spread and heavier tail, leaving PR size as a potential confounder.
We mitigate this issue using a balanced dataset and churn-distribution analysis, but broader benchmarks are needed.

\par\noindent\textbf{Construct validity:}
Precision, recall, and F1-score measure PR-level binary regression detection rather than oracle quality, intent fidelity, or exact bug-instance matching.
Manual inspection shows that 38 of the 49 PRs predicted by \toolname at least partially cover the developer-reported regression.
Baselines show similar limitations but detect fewer positives, with only 6/18, 7/14, and 9/14 cases consistent for Testora, the pure LLM baseline, and Codex, respectively.
Thus, our results should be interpreted as PR-level detection rather than exact localization or full reproduction.
Because the exact training corpora of the evaluated LLMs are
unavailable, we cannot exclude the possibility that some public
GitHub PRs in RegBench-Py were encountered during model pretraining.
\toolname additionally relies on executable pre/post evidence rather than
model-only judgment, but this does not eliminate the potential
training-data leakage threat.

\section{Related Work}
\label{sec:related}

\subsection{Regression Detection and Behavioral Differences}
Regression detection has been extensively studied through regression test selection, prioritization, and minimization~\cite{yoo2012regression, engstrom2010systematic, rothermel2001prioritizing}, which improve testing efficiency but depend on existing test suites and may miss uncovered regressions.
Other work detects behavioral differences across program versions, including differential testing and its variants~\cite{mckeeman1998differential, petsios2017nezha, da2020detecting}, as well as change-aware techniques such as differential assertion checking~\cite{lahiri2013differential}.
However, these approaches mainly expose differences rather than determine whether they are intended.
Commits and PRs have also been used for defect prediction and change risk analysis~\cite{kim2008classifying, kamei2012large}, while recent PR-level regression detection work~\cite{pradel2025testora} still relies on unstructured diffs and lacks explicit behavioral modeling.
PatchGuru~\cite{lepatchguru} is also closely related to ORION in that it reasons about
PR intent and operationalizes intended behavior into executable
assertions paired with concrete inputs. ORION differs in first
constructing a function-indexed structured behavioral oracle that
separately represents preservation requirements, intended evolution,
setup requirements, and control-flow conditions. These oracle items
subsequently guide multiple executable tests and targeted
augmentation. Moreover, PatchGuru's current implementation targets
single-function PRs, whereas ORION supports PRs involving multiple
impacted functions and retains local caller/callee context.
In contrast, \toolname performs intent-aware regression detection by jointly modeling structural changes and expected behavior.

\subsection{Automated Test Generation}
Automated test generation includes random testing~\cite{pacheco2007randoop}, search-based testing~\cite{fraser2011evosuite}, symbolic/concolic execution, and hybrid approaches~\cite{godefroid2008automated, tillmann2008pex}, which improve coverage but are largely agnostic to code-change semantics.
Recent LLM-based techniques further generate unit tests, fuzzing inputs, and bug-reproduction tests~\cite{tufano2020unit, chen2021evaluating, schafer2023empirical, alagarsamy2024a3test, zheng2023codegeex, fakhoury2024llm, wang2024hits, gu2024testart, ryan2024code}, often with feedback loops or search integration~\cite{altmayer2025coverup, deljouyi2025leveraging, chen2025revisit}.
Despite their effectiveness, most methods optimize general coverage or mutation score~\cite{fraser2011evosuite, papadakis2019mutation} and entangle behavioral reasoning with test synthesis, limiting their ability to target regression-relevant behaviors and reducing interpretability.
\toolname instead separates behavioral-oracle synthesis from test generation and explicitly targets regression-inducing conditions.

\subsection{Modeling Code Changes, Intent, and Oracles.}
Understanding code evolution requires reasoning about both structural changes and developer intent.
Prior work mines version histories to predict defects and characterize changes~\cite{kim2008classifying, kim2007predicting, mockus2000predicting}, while empirical studies show that subtle interface changes and dynamic-language behaviors are important regression sources~\cite{brito2020you, kula2018empirical, ge2017refactoring, serbout2023empirical}.
Natural language artifacts have been used to infer intent~\cite{pradel2025testora, di2016would}, but are typically treated as unstructured input.
Meanwhile, the test oracle problem remains fundamental~\cite{6963470, weyuker1982testing}; existing techniques derive oracles from specifications, invariants, metamorphic relations~\cite{chen2020metamorphic, segura2016survey}, or oracle-aware regression test selection~\cite{yu2013oracle}.
These approaches usually assume stable expected behavior and do not explicitly model evolving expectations across versions.
In contrast, \toolname integrates intent and structural signals to construct change-aware oracles that distinguish preserved and changed behaviors.

\subsection{LLM-Based Program Analysis}
LLMs have been widely applied to program analysis tasks~\cite{chen2021evaluating, fan2023large, pearce2025asleep, chen2024survey}, but still struggle with fine-grained structural reasoning and consistency on code changes~\cite{siddiq2024using, cordeiro2025llm}.
Prior work often incorporates lightweight static analysis to guide LLM reasoning, yet few systems explicitly couple static risk signals, structured behavioral oracles, and differential execution for PR-level regression detection.
\toolname adopts this staged design to improve robustness and interpretability without heavy analysis overhead.

\section{Conclusion}
\label{sec:conclusion}
We present \toolname, a PR-level behavioral regression detection framework that combines PR intent modeling, structured oracle synthesis, oracle-guided test generation, and differential execution. We construct RegBench-Py, a benchmark of 154 Python PRs with verified labels. \toolname achieves F1-scores of 0.6125 with DeepSeek-Chat-V3.2 and 0.5578 with Claude-4.5-haiku, outperforming the baselines evaluated on the full benchmark under the corresponding settings. A scope-aware comparison with PatchGuru shows that neither method consistently dominates on the commonly completed single-function PRs, while \toolname supports a broader range of multi-function PRs. \toolname requires approximately 15 minutes per PR, making it practical for asynchronous CI/CD or high-risk PR analysis. In an extended real-world analysis of 1,459 PRs, \toolname identifies 9 reproducible bug candidates, 8 of which have been confirmed by maintainers. Overall, these results demonstrate that structured behavioral oracles provide a practical foundation for intent-aware PR-level regression detection.

\section{Data Availability Statement}
Our source code and experimental data are available at 
\url{https://github.com/wenwen-666/ORION}.

\bibliographystyle{IEEEtran}
\bibliography{sample-base}

\end{document}